\documentclass[conference]{IEEEtran}
\IEEEoverridecommandlockouts
\usepackage{cite}
\usepackage{amsmath,amssymb,amsfonts}
\usepackage{algorithmic}
\usepackage{graphicx}
\usepackage{textcomp}
\usepackage{xcolor}
\usepackage{tcolorbox}
\usepackage{subcaption}
\usepackage{multirow}
\tcbuselibrary{skins}
\def\BibTeX{{\rm B\kern-.05em{\sc i\kern-.025em b}\kern-.08em
    T\kern-.1667em\lower.7ex\hbox{E}\kern-.125emX}}
\begin{document}

\title{LLM-based Multi-Agent System for Simulating Strategic and Goal-Oriented Data Marketplaces
}

\makeatletter
\newcommand{\linebreakand}{
  \end{@IEEEauthorhalign}
  \hfill\mbox{}\par
  \mbox{}\hfill\begin{@IEEEauthorhalign}
}
\makeatother

\author{
\IEEEauthorblockN{Jun Sashihara}
\IEEEauthorblockA{
\textit{The University of Tokyo}\\
Tokyo, Japan\\
sashihara-jun0116@g.ecc.u-tokyo.ac.jp
}
\and
\IEEEauthorblockN{Yukihisa Fujita}
\IEEEauthorblockA{
\textit{Toyota Motor Corporation}\\
Tokyo, Japan\\
yukihisa.fujita.1@toyota.global
}
\and
\IEEEauthorblockN{Kota Nakamura}
\IEEEauthorblockA{
\textit{Toyota Motor Corporation}\\
Tokyo, Japan\\
kota.nakamura.10@toyota.global
}
\linebreakand
\IEEEauthorblockN{Masahiro Kuwahara}
\IEEEauthorblockA{
\textit{Toyota Motor Corporation}\\
Tokyo, Japan\\
masahiro.kuwahara.1@toyota.global
}
\and
\IEEEauthorblockN{Teruaki Hayashi}
\IEEEauthorblockA{
\textit{The University of Tokyo}\\
Tokyo, Japan\\
hayashi@sys.t.u-tokyo.ac.jp
}
}

\maketitle

\begin{abstract}
Data marketplaces, which mediate the purchase and exchange of data from third parties, have attracted growing attention for reducing the cost and effort of data collection while enabling the trading of diverse datasets. However, a systematic understanding of the interactions between market participants, data, and regulations remains limited. To address this gap, we propose a Large Language Model-based Multi-Agent System (LLM-MAS) for data marketplaces. In our framework, buyer and seller agents powered by LLMs operate with explicit objectives and autonomously perform strategic actions, such as—planning, searching, purchasing, pricing, and updating data. These agents can reason about market dynamics, forecast future demand, and adapt their strategies accordingly. Unlike conventional model-based simulations, which are typically constrained to predefined rules, LLM-MAS supports broader and more adaptive behavior selection through natural language reasoning. We evaluated the framework via simulation experiments using three distribution-based metrics: (1) the number of purchases per dataset, (2) the number of purchases per buyer, and (3) the number of repeated purchases of the same dataset. The results demonstrate that LLM-MAS more faithfully reproduces trading patterns observed in real data marketplaces compared to traditional approaches, and further captures the emergence and evolution of market trends.
\end{abstract}


\section{Introduction}
Data have emerged as a tradable economic resource, and data marketplaces that mediate the purchase and exchange of datasets from third parties have rapidly expanded \cite{rusche2019data}. These marketplaces streamline data collection that previously required substantial cost and effort, while also providing organizations and researchers with access to diverse, high-quality datasets. As a result, they are increasingly recognized as critical infrastructures that accelerate innovation based on data that were closed within individual organizations \cite{balazinska2011data}.

Despite this progress, our understanding of how interactions among market participants, data, and regulations shape market dynamics remains limited. Smooth and efficient data transactions require well-designed and robust data marketplaces \cite{liang2018survey}. The importance of simulation-based approaches to bridge the gap between market design theory and practice has been emphasized \cite{fernandez2020data}. However, most existing simulations rely on model-based approaches with simple, predefined trading criteria. Such approaches are insufficient to capture essential real-world characteristics, including the heterogeneous objectives of participants, their strategic decision-making processes, and the emergence of domain-specific or temporal trends. Consequently, these models cannot adequately reproduce or analyze key phenomena such as price formation, long-term demand fluctuations, repeated data usage, and value updates.

To address these challenges, we propose a Large Language Model-based Multi-Agent System (LLM-MAS) for simulating data marketplaces. Buyer and seller agents in this system are endowed with explicit objectives and can reason in natural language to take strategic actions such as planning, searching, purchasing, pricing, and updating data. This design enables agents to reproduce complex decision-making, forecast demand, and adapt strategies to evolving market conditions. In contrast to conventional rule-based simulations, our framework supports broader, adaptive, and goal-oriented behavior selection through natural language reasoning. Figure  \ref{fig:overall_architecture} shows the overall architecture of our proposed system. Based on these motivations, this study makes the following contributions:
\begin{enumerate}
\item Market-level simulation architecture: We design an LLM-MAS that captures both macro-level market structures and long-term dynamics such as demand cycles and trend emergence.
\item Agent-level behavioral modeling: We implement detailed buyer and seller models capable of goal-oriented actions and natural language reasoning, capturing diverse decision-making and interactions.
\item Data market evaluation metrics: We introduce real-world evaluation metrics that connect micro-level agent behavior with macro-level market outcomes, validated against transaction records from an actual data marketplace.
\end{enumerate}

\begin{figure*}
    \centering
    \includegraphics[width=0.9\linewidth]{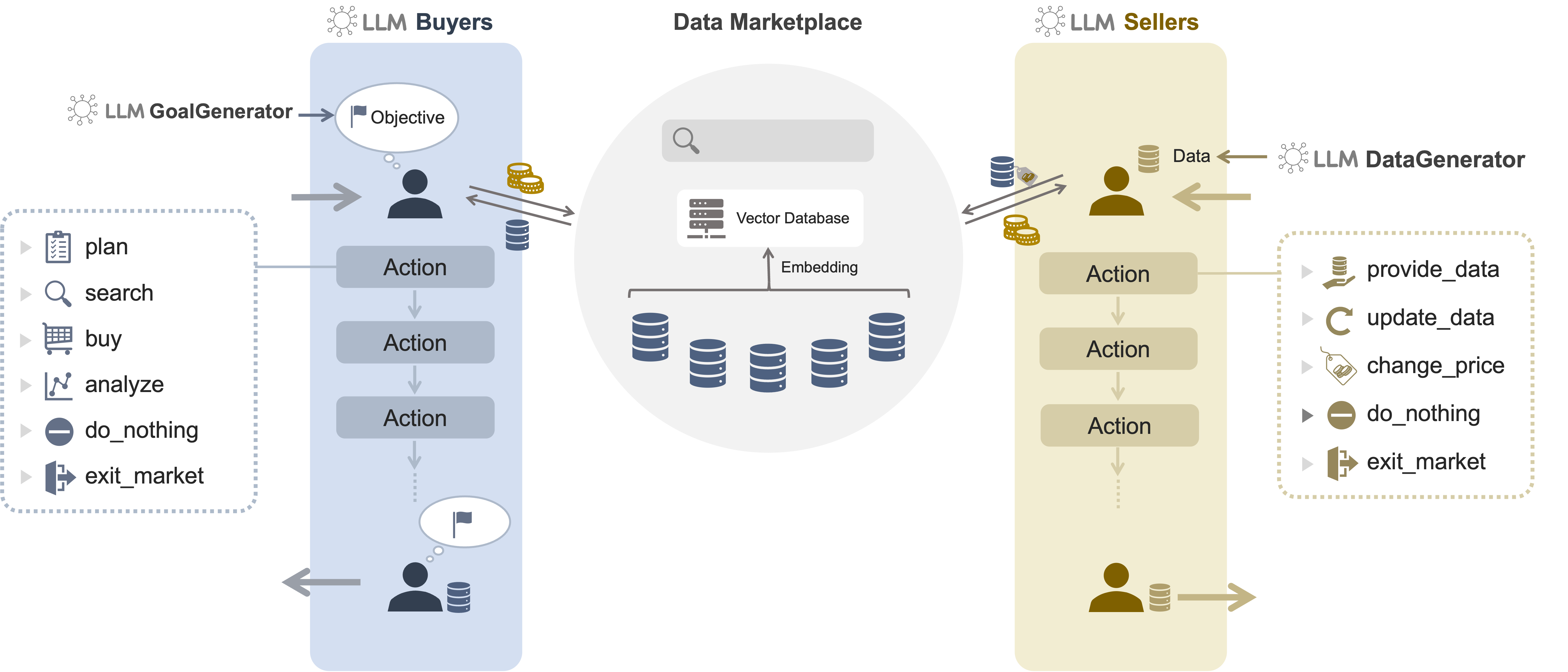}
    \caption{Overall architecture of the proposed LLM-MAS for data marketplaces. The system consists of seller agents providing and updating data, and buyer agents pursuing goal-oriented acquisitions, a GoalGenerator for assigning analytical objectives, and a DataGenerator for producing metadata. Embedded metadata in a vector database supports cosine similarity-based search, reproducing the strategic behaviors, demand fluctuations, and trend emergence observed in real-world data marketplaces.}
    \label{fig:overall_architecture}
\end{figure*}

These contributions collectively provide a structural perspective on data marketplace dynamics: how micro-level strategic interactions among heterogeneous agents give rise to macro-level phenomena. Such a perspective lays the foundation for future research on market and institutional design in which, simulation-based insights can inform policy-making, pricing mechanisms, and data management strategies.

\section{Related Work}
Data marketplaces, which include data suppliers, buyers, and platform operators \cite{stahl2016classification}, serve as a medium for the exchange and monetization of data across various organizations, gaining significant attention as a vital infrastructure in the data economy \cite{schomm2013marketplaces, spiekermann2019data}. However, the range of phenomena that can be directly observed in these markets remains limited \cite{abbas2021business}. To address this limitation, Fernandez et al. \cite{fernandez2020data} emphasized the importance of employing simulation-based approaches to study data markets, arguing that model-based simulations are effective tools for analyzing market structures and trading mechanisms. Building on this idea, prior research has created networks in which data are represented as nodes, examined their structural properties \cite{hayashi2020understanding}, conducted a scenario analysis of data purchasing behaviors \cite{hayashi2022models}, and proposed simulations in which buyer agents perform random walks on a data network to evaluate the impact of purchasing strategies \cite{sashihara2024impact}.

However, these approaches face two main obstacles. First, model-based techniques rely extensively on established behavioral rules, making it difficult to understand complex or adaptive decision-making processes \cite{ijcai2024p874}. Second, while data transactions are generally motivated by clear individual goals, these methods assume that all agents are homogeneous, overlooking the various reasons for participating in the market \cite{hayashi2022models}. Moreover, in contrast to financial markets where stylized facts and established simulation benchmarks are present \cite{chiarella2002simulation}, data markets do not have standardized market features or evaluation criteria. This lack of standardization impedes the use of current simulation methods in the practical development of real-world markets.

The advent of LLMs has presented novel opportunities to address these constraints. Owing to their sophisticated capabilities in natural language comprehension and reasoning, LLMs facilitate innovative simulation frameworks \cite{wang2024survey}. For example, Generative Agents \cite{park2023generative} introduced a framework in which agents powered by LLMs interact and operate over extended periods. ReAct \cite{yao2023react} merges reasoning with action generation, enabling agents to adapt plans dynamically and engage with their surroundings to acquire new information. These frameworks have demonstrated that LLM-based agents are highly proficient in executing complex tasks and exhibiting emergent behaviors \cite{bo2024reflective}.

Building on this foundation, recent research has utilized LLM-MAS in social simulations. For instance, TwinMarket \cite{yang2025twinmarket} has shown its capability in examining emergent phenomena that connect individual investor behaviors at the micro-level to broader financial market trends. AgentSociety \cite{piao2025agentsociety} incorporated psychological theories and a three-layer social environment into an LLM-MAS framework to investigate societal challenges such as political polarization and the implementation of universal basic income.

Our research leverages recent advancements by adapting LLM-MAS to data marketplaces. We create a market setting in which LLM-based agents function as data buyers and sellers, capturing diverse motivations and complex dynamics that traditional model-based simulations cannot address. In addition, we derive market-specific attributes from real transaction data and assess how effectively these characteristics manifest in simulation outcomes to establish a more authentic simulation framework for market and institutional design.

\section{Our LLM-MAS Architecture}
\subsection{Data and Metadata}
Tabular datasets are among the most widely used and practically manageable data formats. In our simulation, we assume that all data traded are tabular. In real marketplaces, buyers generally cannot inspect raw data before purchase; instead, their decisions rely primarily on metadata, which describes attributes and structure rather than raw content. Metadata enables potential users to assess whether a dataset aligns with their analytical objectives. In our framework, all the information exchanged during transactions is represented as metadata.

Metadata are generated by the DataGenerator agent.
Figure \ref{fig:prompt_for_datagenerator} shows the prompt provided to the DataGenerator to produce this metadata, which consists of six fields (Table \ref{tab:metadata_generated_by_datagenerator}): \textit{data\_name}, \textit{description}, \textit{columns}, \textit{tags}, \textit{data\_price}, and \textit{update\_frequency}. The \textit{description} provides a natural language summary of the content of the dataset. For example: 
\begin{tcolorbox}[
    colframe=white, 
    colback=white, 
    before skip = 0pt,
    after skip = 0pt,
    fontupper=\ttfamily\small
    ]
        This dataset contains detailed shot chart analytics of professional basketball players worldwide during the 2024 season, capturing shot locations, shot types, outcomes, and player efficiency metrics.
    \end{tcolorbox}

Column information is included to describe the table schema, which is essential for assessing the dataset suitability and potential interoperability with other datasets. Prior studies have highlighted the importance of leveraging column information for dataset similarity calculations \cite{sakaji2020verification}.

In real-world marketplaces, both public (often free) and private (paid) datasets coexist. Operators frequently seed the platform with high-quality public data to attract users and later expand with premium private offerings. Although our framework could model the dynamic ratio of public to private data, we simplified this aspect. Prices are autonomously determined by the DataGenerator. 

Among metadata fields, \textit{update\_frequency} is particularly important. It reflects how often data is refreshed and categorized into four levels: \textit{static}, \textit{low}, \textit{medium}, and \textit{high}. Static data (e.g., finalized financial statements and genomic sequences) are rarely updated, whereas dynamic data (e.g., traffic records, sensor streams, and stock prices) are frequently updated.  Seller agents consider both the update frequency and sales history when deciding whether to update data. All generated metadata were embedded using OpenAI's text-embedding-3-large model and stored in a vector database. This enables efficient similarity-based retrieval, supporting buyers' search actions during the simulation.

\begin{table}[]
    \centering
    \caption{Metadata generated by DataGenerator}
    \scriptsize
    \renewcommand{\arraystretch}{1.4}
    \begin{tabular}{ll}
        \hline
        \textbf{Field} & \textbf{Description} \\
        \hline
        \textit{data\_name} & Name of the data \\
        \textit{description} & Natural language summary of the dataset's content \\
        \textit{columns} & List of column name \\
        \textit{tags} & Keywords describing topical domains \\
        \textit{data\_price} & Seller-specified price \\
        \textit{update\_frequency} & Refresh frequency (\textit{static}, \textit{low}, \textit{medium}, \textit{high}) \\
        \hline
    \end{tabular}
    \label{tab:metadata_generated_by_datagenerator}
\end{table}

\subsection{Models of Buyer and Seller}
Buyer and seller agents represent the primary market participants (Fig.~\ref{fig:overall_architecture}). Their objectives and available actions are summarized in Table \ref{tab:seller_and_buyer_in_llm-mas_for_data_marketplaces}. Sellers aim to maximize revenue through data provision and pricing, while buyers pursue specific analytical objectives through strategic purchasing and analysis by considering their budgets and purchase histories.

\begin{table*}[]
    \centering
    \caption{Seller and buyer in LLM-MAS for data marketplaces}
    \scriptsize
    \renewcommand{\arraystretch}{1.4}
    \begin{tabular}{lll}
        \hline
        \textbf{Agent} & \textbf{Objective} & \textbf{Available Actions} \\
        \hline
        Seller & Maximize profit by providing and updating datasets and setting prices & \textit{update\_data}, \textit{change\_price}, \textit{provide\_data}, \textit{do\_nothing}, \textit{exit\_market} \\
        Buyer & Achieve a specific analytical goal through strategic purchasing and data analysis & \textit{plan}, \textit{search}, \textit{buy}, \textit{analyze}, \textit{do\_nothing}, \textit{exit\_market}\\
        \hline
    \end{tabular}
    \label{tab:seller_and_buyer_in_llm-mas_for_data_marketplaces}
\end{table*}

\noindent \textbf{Seller Agent}

Each seller agent seeks to maximize profit by managing its data portfolio. At every simulation step, the seller selects one of five possible actions guided by prompts shown in Fig. \ref{fig:system_prompt_for_seller}:

\textbf{\textit{update\_data}}: This action updates the version of the data already provided by the seller; for example, by adding new information or revising existing content. The purpose is to enhance the value of the data and increase its attractiveness to potential buyers. In domains such as finance, where ongoing daily analysis is common, the freshness of the data is critical. The seller evaluates whether the data requires updating based on its metadata field \textit{update\_frequency} and its own action history, and chooses \textit{update\_data} when necessary.
    
\textbf{\textit{change\_price}}: When this action is selected, the seller adjusts the price of the data. In many cases, this involves lowering the price if sales are poor. The size of the price adjustment is not constrained, allowing the agent to autonomously determine the appropriate magnitude.

\textbf{\textit{provide\_data}}: This action introduces entirely new data to the marketplace in addition to the seller's existing offerings. Although creating new data incurs costs, the seller may decide to provide new data when doing so is expected to significantly influence market conditions.

\textbf{\textit{do\_nothing}}: In real-world data marketplaces, sellers generally act less frequently than buyers because modifying or creating datasets entails nontrivial costs. As a result, the default long-term strategy for a seller is to remain inactive unless there is a clear reason to act. The system prompt explicitly instructs sellers to select \textit{do\_nothing} unless special circumstances warrant another action.

\textbf{\textit{exit\_market}}: Finally, if a seller determines that its datasets are rarely sold and no future profit can be expected, it may withdraw from the marketplace by removing its datasets.

\vskip0.5\baselineskip
\noindent \textbf{Buyer Agent}

Each buyer agent is initialized with a unique analytical objective expressed in natural language. This reflects the diversity of real-world motivations; for instance, a financial institution may analyze customer transaction data for credit risk modeling, while an urban planning office may investigate traffic congestion patterns to optimize infrastructure. Traditional model-based approaches assume homogeneous buyer objectives. However, our framework captures heterogeneity by generating goals in natural language using the GoalGenerator agent. Two types of prompts were used as inputs (Figs.~\ref{fig:prompt_with_trends_for_goalgenerator} and \ref{fig:prompt_without_trends_for_goalgenerator}).

\textbf{Initial buyers}: At the start of the simulation, buyers are randomly assigned one of 10 predefined fields (cybersecurity, education, environmental science, finance, healthcare, manufacturing, marketing, social media, sports, and urban planning), chosen from demands in both commercial data marketplaces and public open-data portals. Goals were generated without reference to trends. For example:

\begin{tcolorbox}[
    colframe=white, 
    colback=white, 
    before skip = 0pt,
    after skip = 0pt,
    fontupper=\ttfamily\small
    ]
        Analyze player performance metrics and game statistics from professional basketball leagues to identify key factors influencing team winning percentages and build a predictive model for upcoming game outcomes.
    \end{tcolorbox}

\textbf{New entrants}: In contrast, buyers entering later may be trend-sensitive, i.e., the objectives of participants entering the market can be influenced by ongoing market trends. For example, if competitors are performing certain analyses, failing to conduct similar analyses internally may weaken a company's competitive advantage. To capture such dynamics, each new buyer's goal is generated by feeding GoalGenerator a prompt that incorporates trends with a probability of 0.5. At each step, the buyer agent selects one of six possible actions (Table \ref{tab:seller_and_buyer_in_llm-mas_for_data_marketplaces}).

\textbf{\textit{plan}}: Develop a strategy by specifying the required datasets and analysis methods, typically chosen at the initial stage of market participation or after purchasing and analyzing some datasets. During planning, the buyer establishes a long-term strategy that includes the types of data required and the analysis procedures needed. For example:
    \begin{tcolorbox}[
    colframe=white, 
    colback=white, 
    before skip = 0pt,
    after skip = 0pt,
    fontupper=\ttfamily\small
    ]
        To analyze player performance metrics and game statistics from professional basketball leagues and identify key factors influencing team winning percentages, I will first plan to search for comprehensive datasets including player stats, game stats, and team performance data.
    \end{tcolorbox}

\textbf{\textit{search}}: Issue natural language queries to retrieve the related dataset. In real-world data marketplaces, it is impractical to directly review all available data. Instead, as in actual platforms, users typically issue queries or apply filters to identify relevant candidates. When this action is selected, the buyer agent composes a query and performs cosine similarity-based retrieval over the metadata. The top 20 results are provided as purchase candidates from which the buyer makes a final decision using \textit{buy}. The query example is as follows:

\begin{tcolorbox}[
    colframe=white, 
    colback=white, 
    before skip = 0pt,
    after skip = 0pt,
    fontupper=\ttfamily\small
    ]
        professional basketball player performance, game statistics, team winning percentages, seasonal data
    \end{tcolorbox}

\textbf{\textit{buy}}: Purchases data based on the current plan.

\textbf{\textit{analyze}}: Derive insights from purchased datasets.

\textbf{\textit{do\_nothing}}: Wait when suitable data is unavailable or market conditions are unfavorable.

\textbf{\textit{exit\_market}}: Withdraw if goals cannot be achieved or objectives have already been met. 

The fundamental setup for each buyer agent is outlined as a system prompt (Fig. \ref{fig:system_prompt_for_buyer}). At each step, the buyer also receives a user prompt detailing its analytical goals, budget, possessed datasets, action history (including the plan), and the latest search outcomes (Fig. \ref{fig:user_prompt_for_buyer}). By integrating this long-term strategy with its current state, the agent can assess its circumstances and determine the best action for that step.

The buyer's choice to leave the market is guided by a combination of an LLM-based autonomous \textit{exit\_market} decision and two specific rule-based criteria: (1) the \textit{analyze} to \textit{buy} ratio surpasses two, and (2) \textit{plan} is chosen more than seven times in a row. The first condition implies that sufficient data have been gathered, making further purchases improbable. The second condition indicates that the buyer has limited viable options in the current market setting. By merging natural language reasoning with these rule-based conditions, the framework implements the decision for buyer withdrawal.

\begin{table*}[]
    \centering
    \caption{Summary of the five macro-level evaluation metrics used to assess simulation results against real data}
    \scriptsize
    \renewcommand{\arraystretch}{1.4}
    \begin{tabular}{p{1.3cm} p{5cm} p{10cm}}
        \hline
        \textbf{Category} & \textbf{Metric} & \textbf{Description} \\
        \hline
        \multirow{3}{*}{Structural}
        & Frequency of data purchases & Frequency distribution of how many times each dataset is purchased (expected to follow a long-tail). \\
        & Frequency of purchases per buyer & Distribution of purchase counts per buyer measuring buyer activity level and market engagement. \\
        & Frequency of repeated purchases of the same data & Counts how often the same buyer purchases the same data repeatedly reflecting continuous demand. \\
        \hline
        Network 
        & Degree distribution of buyer–seller bipartite graph & A bipartite network and examines the degree distribution, capturing scale-free characteristics. \\
        \hline
        Dynamic
        & Autocorrelation of transaction volume over time & Quantifies temporal persistence of trading activity. \\
        \hline
    \end{tabular}
    \label{tab:summary_of_the_five_evaluation_metrics}
\end{table*}

\section{Experiment settings and Evaluation}
\subsection{Simulation Process}
The LLM-MAS simulation process is shown in Fig.~\ref{fig:simulation_process}. This simulation advances in distinct steps, in which the buyer and seller agents engage in natural language reasoning to strategize, conduct trades, and update datasets. During the initialization phase, a predetermined number of buyers and sellers is created based on the parameter settings. Each seller is linked to a single dataset upon creation, making the initial count of tradable datasets equal to the number of sellers. Metadata is produced by the DataGenerator and stored in a vector database. Buyers are provided with their analytical objectives from the GoalGenerator and are given a budget that is randomly selected from predefined levels.

The primary simulation loop continues until it reaches the set maximum number of steps or until all agents have left the market. During each step, the simulation begins with the dynamic entry of agents. Depending on the recent transaction volume, new buyers and sellers may join the market. The number of new agents is determined by a sigmoid function, which simulates the idea that higher transaction volumes draw more market participants. For new buyers, there is a 0.5 probability that their goals will reflect market trends based on the most popular selling datasets. In the seller phase, each active seller chooses an action from options, guided by their sales history and metadata. Any newly provided or updated datasets are added to the shared vector store.

The buyer phase follows, during which each active buyer selects an action. If a buyer opts for \textit{search}, a similarity-based retrieval from the vector store suggests potential datasets for purchase. Finally, in the transaction phase, purchase requests from buyers are matched with sellers' offerings. If the buyer's budget allows, the transaction is completed successfully, budgets are adjusted, and the seller's revenue and sales history are updated.

\begin{figure}
    \centering
    \includegraphics[width=0.5\linewidth]{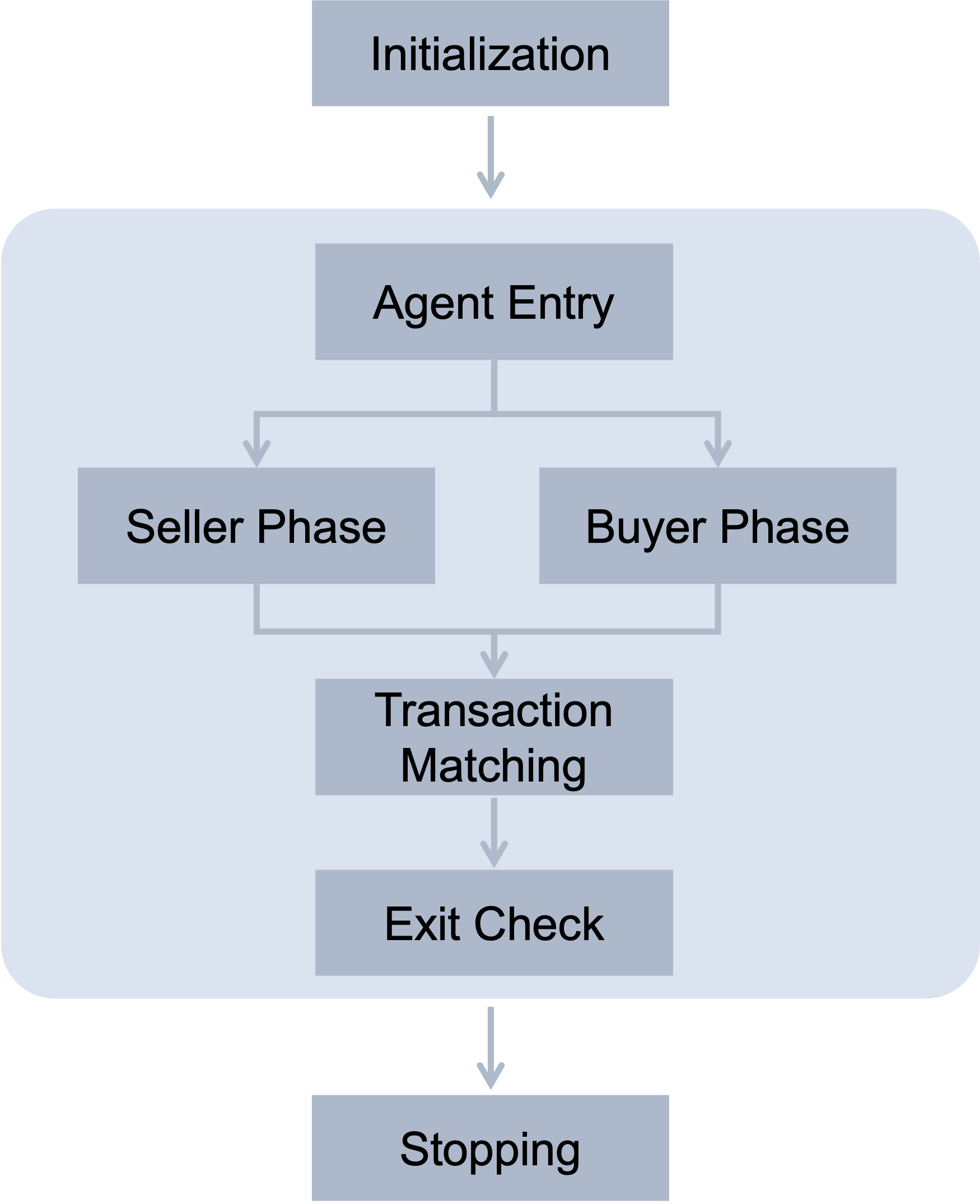}
    \caption{Simulation process}
    \label{fig:simulation_process}
\end{figure}

\subsection{Parameters}
The simulation begins with 10 buyers and 5 sellers, with each seller initially possessing one dataset. As a result, five datasets were available for trade at the outset. Each buyer's starting budget is randomly selected from the set \{1,000, 10,000, and 100,000\} units. The simulation can run for up to 40 steps unless all participants leave the market sooner. The entry of new agents is regulated by a sigmoid function:
\begin{equation}
f(x) = \frac{L}{1 + e^{-k(x - x_0)}},
\end{equation}
where \(x\) represents the recent transaction volume; \(L\) is the maximum number of entrants per step, set at 50; \(k\) is the growth rate, set at 0.03; and \(x_0\) is the inflection point, set at 100. For new buyers, the GoalGenerator formulates objectives that reflect current market trends with a probability of 0.5. Sellers will exit if their datasets remain unsold for 10 consecutive steps. Buyers will leave the market if the ratio of \textit{analyze} to \textit{buy} actions surpasses two, or if the \textit{plan} action is chosen more than seven times in a row.

\subsection{Metrics and Benchmarks}
\noindent \textbf{Data transaction dataset}

As data markets are still in their infancy, the availability of public transaction histories is limited. We utilized data from Ocean Protocol, a decentralized data marketplace framework that links data providers and consumers through blockchain technology. Within Ocean Protocol, data providers create DataNFTs to signify ownership of data assets and exchange Datatokens to provide usage rights. For this research, we utilized Ocean Protocol's Subgraph to collect events related to Datatoken transactions. The dataset obtained includes unique identifiers (\textit{order\_id}, \textit{transaction\_hash}), details about the traded Datatoken (\textit{datatoken\_address}, etc.), consumer details (\textit{payer\_address}), transaction volume (\textit{amount}), price (\textit{price}), and timestamp (\textit{timestamp}). Our analysis covers May 30, 2022, to June 20, 2025, during which we collected a total of 6,826 transaction records.

To evaluate how closely our LLM-MAS reproduced the structural and temporal properties of real-world data markets, we defined five evaluation metrics grouped into structural, network, and dynamic categories summarized in Table~\ref{tab:summary_of_the_five_evaluation_metrics}.

\vskip0.5\baselineskip
\noindent \textbf{Structural Metrics: Distributions of Data Transactions}

To encapsulate the essential characteristics of data marketplaces, we identified three principal power-law distributions.

Frequency of data purchases: This metric illustrates the distribution of how often each dataset is bought over the observation period. It assesses the concentration of demand and the popularity of datasets. Similar to many real-world markets, a long-tail effect is expected, in which a few highly popular datasets dominate the majority of transactions.

Frequency of purchases per buyer: This metric examines the distribution of the number of purchases made by each buyer. It indicates the level of buyer activity and the degree of market participation.

Frequency of repeated purchases of the same data: This metric tracks how frequently individual buyers purchase the same dataset multiple times. Given that datasets can be updated and easily replicated, they are often repeatedly bought and reused. Thus, this measure offers insights into the ongoing demand and usage trends driven by dataset updates.

\begin{figure*}
    \centering
    \begin{subfigure}{0.3\textwidth}
        \includegraphics[width=\textwidth]{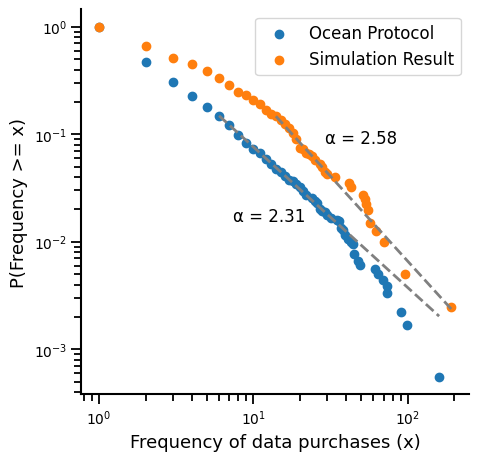}
        \caption{Number of purchases per data}
        \label{fig:number_of_purchases_per_data}
    \end{subfigure}
    \hfill
    \begin{subfigure}{0.3\textwidth}
        \includegraphics[width=\textwidth]{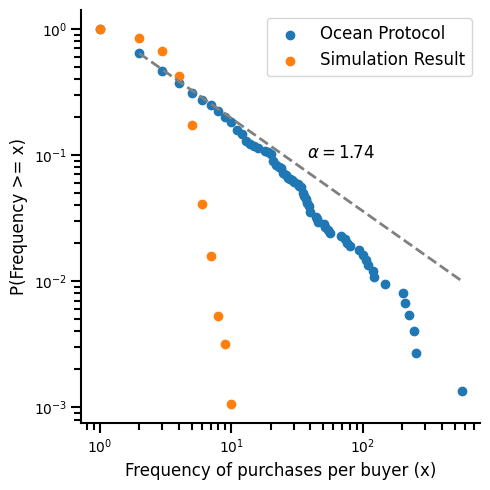}
        \caption{Number of purchases per buyer}
        \label{fig:number_of_purchases_per_buyer}
    \end{subfigure}
    \hfill
    \begin{subfigure}{0.3\textwidth}
        \includegraphics[width=\textwidth]{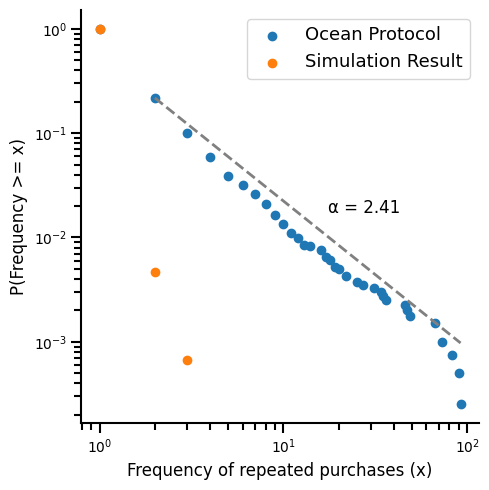}
        \caption{Number of repeated transactions}
        \label{fig:number_of_repeated_purchases}
    \end{subfigure}
    \caption{Three distributions of data transactions in a real-world data market and simulation results}
    \label{fig:three_distribution_of_data_transactions}
\end{figure*}

\vskip0.5\baselineskip
\noindent \textbf{Network Metrics: Transaction Network Structure}

Although structural metrics can capture the overall distribution within a market, they fail to illustrate the connections between participants. To describe these relational patterns, we represent all buyer-seller interactions as a bipartite graph, where each node and edge signifies a participant and a completed transaction, respectively. We then analyzed the degree distribution of this network to identify the presence of hub nodes and the concentration of transactions. This network-level examination enhances structural metrics by elucidating the extent of connections among participants and determining whether the marketplace displays a scale-free topology—a characteristic often observed in complex economic systems.

\vskip0.5\baselineskip
\noindent \textbf{Dynamic Metrics: Temporal Market Dynamics}

We performed a time-series analysis of market dynamics to assess the effectiveness of the proposed simulation in replicating the continuity of trading activity. We utilized the autocorrelation coefficient of transaction volume over time as the primary dynamic measure. A higher coefficient suggests that once trading becomes active, increased transaction volumes are likely to persist across subsequent simulation steps, thereby reflecting the temporal dependence of market activity.

\section{Result and Discussion}
\subsection{Macro-Level Evaluation}
Over the course of the 40-step simulation, the data marketplace expanded to a total of 1,149 buyers and 1,144 sellers, generating 3,014 data transactions (Fig. \ref{fig:number_of_transations}). Among all participants, 950 buyers made at least one purchase, while 365 sellers successfully sold at least once. In total, 401 unique datasets were traded during the simulation period (Fig. \ref{fig:cumulative_count}).

\vskip0.5\baselineskip
\noindent \textbf{Structural Metrics: Distributions of Data Transactions}

\textbf{Purchases per data}: The distribution of transactions per data exhibited a clear long-tail. The minimum was 1, the maximum was 192, and the mean was 7.52. Fitting this distribution to a power-law yielded an estimated exponent of 2.58, a value that closely aligns with that observed in real-world data marketplaces. The Kolmogorov-Smirnov (KS) distance was estimated at 0.067, indicating a good fit to the power-law model. An exponent within the 2–3 range suggests a scale-free structure in which a small number of highly important datasets dominate overall trading. For comparison, Ocean Protocol’s transaction data exhibit a power-law exponent of 2.30 and KS distance of 0.0259, suggesting that our simulation successfully reproduced a key feature of real-world data
marketplaces (Fig.~\ref{fig:number_of_purchases_per_data}).

\textbf{Purchases per buyer}: The minimum number of purchases was 1, the maximum was 10, the mean was 3.17, and the mode was 4. Fitting this distribution to a power-law yielded an exponent of 11.84, which differs significantly from the corresponding value estimated from Ocean Protocol's real data (1.06) (Fig.~\ref{fig:number_of_purchases_per_buyer}). The simulated distribution clearly followed a Poisson form rather than a power-law. This discrepancy indicates that our simulation does not fully reproduce the empirical distribution of buyer purchase frequencies observed in actual data markets. Specifically, our simulation exhibited a high proportion of buyers making only a few purchases (2-5 times), while extremely active buyers (heavy users) were absent. In contrast, Ocean Protocol data include buyers with more than 500 transactions. Because our simulation was limited to 40 steps, reproducing hundreds of purchases per buyer was inherently difficult. Nevertheless, the fact that the maximum number of purchases was only 10 reveals a design limitation. One possible reason is that, in our model, each buyer must sequentially select from six actions: \textit{plan}, \textit{search}, \textit{buy}, \textit{analyze}, \textit{do\_nothing}, and \textit{exit\_market}. This stepwise decision process constrains the frequency of \textit{buy} actions, thereby reducing the total number of transactions compared with real-world data marketplaces. A longer-term simulation combined with dynamic prompt refinement could potentially capture more complex purchasing behaviors, such as the emergence of heavy users and long-term buying cycles.

\textbf{Repeat purchases for the same data}:
Among all observed transactions, 2,984 pairs involved only a single purchase, 12 pairs involved two purchases, and 2 pairs involved three purchases, indicating that the vast majority of buyer-dataset relationships ended after a single transaction (Fig.~\ref{fig:number_of_repeated_purchases}). All datasets that were purchased multiple times by the same buyer exhibited \textit{update\_frequency} levels of medium or high, characteristic of dynamic data (e.g., GlobalFootballPlayerPerformanceMetrics2024). This pattern suggests that repeated purchases were likely motivated by content changes associated with dataset version updates. In contrast, regular repeat purchases are common in real data marketplaces, yet the frequency of such behavior in our simulation was noticeably lower. One possible explanation is that buyers did not fully recognize or exploit the update frequency of datasets. For example, if buyers were able to identify through \textit{search} datasets that matched their analytical objectives and had high update frequencies, and then plan for periodic re-purchases within \textit{plan}, the simulation might better reproduce the continuous purchasing behavior observed in real markets.

\begin{figure*}
    \centering
    \begin{subfigure}{0.3\textwidth}
        \includegraphics[width=\textwidth]{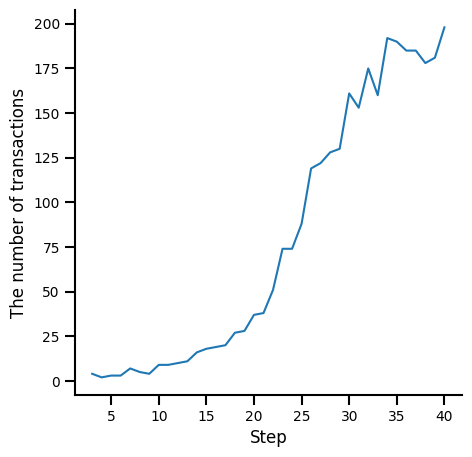}
        \caption{Number of transactions}
        \label{fig:number_of_transations}
    \end{subfigure}
    \hfill
    \begin{subfigure}{0.3\textwidth}
        \includegraphics[width=\textwidth]{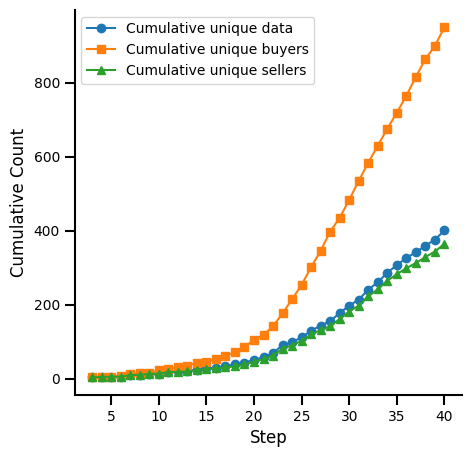}
        \caption{Cumulative count}
        \label{fig:cumulative_count}
    \end{subfigure}
    \hfill
    \begin{subfigure}{0.3\textwidth}
        \includegraphics[width=\textwidth]{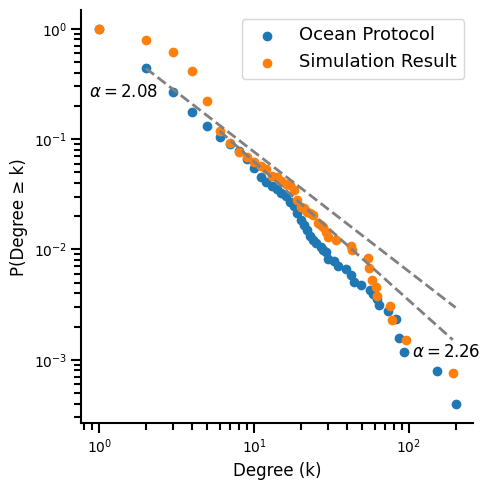}
        \caption{Degree distribution}
        \label{fig:degree_distribution}
    \end{subfigure}
    \caption{Number of data transactions, cumulative count of simulation components, and degree distribution of buyer-seller network.}
    \label{fig:market_wide_growth}
\end{figure*}

\begin{figure}
    \centering
    \includegraphics[width=1.0\linewidth]{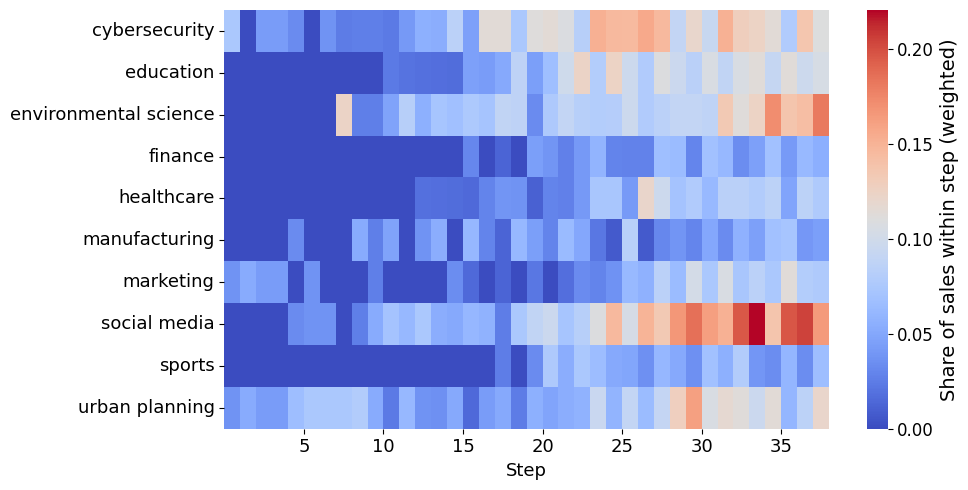}
    \caption{Heatmap for trend}
    \label{fig:heatmap_for_trend}
\end{figure}

\vskip0.5\baselineskip
\noindent \textbf{Network Metric: Buyer–Seller Transaction Graph}

To evaluate the structural properties of the market, we represented the transaction history as a buyer-seller bipartite graph and analyzed its network characteristics. Based on the simulation results, the network exhibited an average degree of 4.48. Fitting the degree distribution to the power-law yielded an estimated exponent of 2.26, indicating a scale-free structure (Fig.~\ref{fig:degree_distribution}). For comparison, the analysis of Ocean Protocol's real transaction data produced an average degree of 3.02 and a power-law exponent of 2.08, suggesting that our simulation successfully reproduced the structural features of buyer-seller relationships observed in real data marketplaces. The complementary cumulative distribution function of the node degree provides visual confirmation of this scale-free property. The highest degree nodes were all sellers. The most connected seller node had 192 links and owned the data ``UrbanAirQualityAndMeteorologicalInterrelationship2024", which recorded the highest transaction count in the entire simulation.

\vskip0.5\baselineskip
\noindent \textbf{Dynamic Metrics: Temporal Market Dynamics}

To evaluate the temporal persistence of market activity, we calculated the autocorrelation coefficient of the transaction count over time. In the simulation, the autocorrelation coefficient of total market transactions was 0.939, indicating strong persistence. Once market activity became active, a high transaction volume tended to be maintained in subsequent steps. In contrast, the same metric computed from real-world transaction data of Ocean Protocol was 0.516, suggesting that while real data markets also exhibit periods of sustained activity, their persistence is considerably weaker than that observed in the simulation.

The excessively high autocorrelation coefficient likely originates from the participant-entry mechanism. In our model, steps with a large number of transactions attract more new participants, many of whom begin by selecting \textit{plan} and typically do not make their first \textit{buy} decision until three or more steps later. As a result, once a period of heavy trading occurs, the market experiences a chain reaction of new entries and subsequent purchases over several steps, making it less likely for the transaction volume to decline. This cascading demand growth is considered a key driver of the high autocorrelation coefficient observed in the simulation.

\subsection{Macro-Level Analysis: Formation of market trends}
In our simulation, we partially incorporated a mechanism in which new market participants considered existing market trends. This design allowed us to observe the emergence and evolution of field-specific demand trends over time. For the analysis, all datasets traded at each step were classified into 10 predefined fields based on the domains assigned at the time of data generation. At the start of the simulation, the interests of participants were evenly distributed across fields, and the market began in a state of diverse but sparse demand. Figure \ref{fig:heatmap_for_trend} shows a heatmap with fields and simulation steps on the vertical and horizontal axes, respectively. The color intensity of each cell represents the field-specific share of transactions within each step, weighted by the total transaction volume of each step.

This visualization reveals the temporal dynamics. During the initial phase, transactions were almost uniformly distributed across the fields. At step 20, short-lived surges of concentrated demand appeared in fields such as cybersecurity. After step 25, cybersecurity and social media emerged as the dominant fields driving the market. Toward the end of the simulation, social media demand continued to grow, while interest in cybersecurity declined, giving way to a late-stage rise in environmental science. These patterns of trend emergence and subsequent shifts mirror the phenomena found in real data marketplaces, where general-purpose markets coexist with niche markets focused on specialized domains. The ability of our simulation to reproduce such dynamics provides useful insights into market design, including strategies to either stimulate toward specific fields or sustain diversified, stable demand over time.

\begin{table*}[]
    \centering
    \caption{Macro-level results, reproduction level relative to real data, and their micro-level causal mechanisms}
    \scriptsize
    \renewcommand{\arraystretch}{1.4}
    \begin{tabular}{p{5cm} p{2cm} p{10cm}}
        \hline
        \textbf{Macro-level metrics} & \textbf{Reproduction} & \textbf{Underlying micro-level behavior} \\
        \hline
        Frequency of data purchases  & Good & Early sellers actively \textit{provide\_data} and \textit{update\_data} creating a skewed supply. \\
        Frequency of purchases per buyer & Poor & Buyers spend early steps on \textit{plan} and \textit{search}, with few \textit{buy} actions. \\
        Frequency of repeated purchases of the same data & Poor & Most buyers may not incorporate dataset \textit{update\_frequency} into their \textit{plan}. \\
        \hline
        Degree distribution of buyer–seller bipartite graph & Good & High-quality sellers become hubs through early \textit{provide\_data} and subsequent \textit{do\_nothing}. \\
        \hline
        Autocorrelation of transaction volume over time & Overestimated & Seller \textit{do\_nothing} stabilizes prices and supply; buyers respond with persistent \textit{buy} or \textit{analyze}. \\
        \hline
    \end{tabular}
    \label{tab:macro_micro_reproduction}
\end{table*}

\subsection{Micro-level Behavioral Analysis}
\begin{figure}
    \centering
    \begin{subfigure}{0.24\textwidth}
        \includegraphics[width=\textwidth]{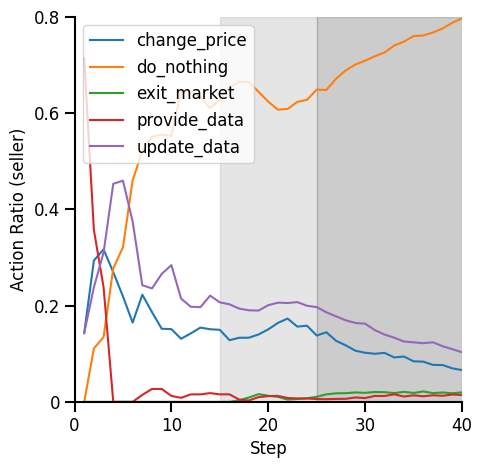}
        \caption{Action Ratio (Seller)}
        \label{fig:action_ratio_seller}
    \end{subfigure}
    \hfill
    \begin{subfigure}{0.24\textwidth}
        \includegraphics[width=\textwidth]{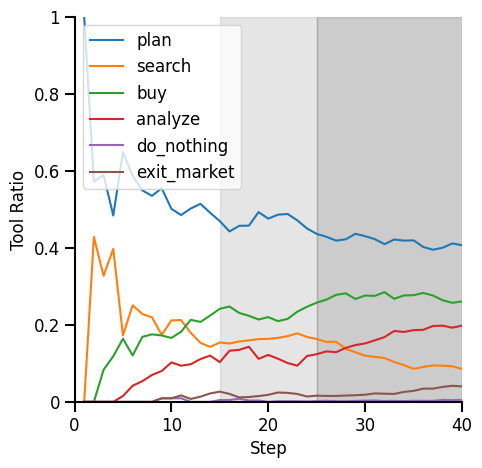}
        \caption{Action Ration (Buyer)}
        \label{fig: action_ration_buyer}
    \end{subfigure}
    \caption{Time-evolution of the seller and buyer action ratios smoothed with a three-step moving average. The light gray shaded area (steps 15–25) and the darker gray shaded area (steps 25–40).}
    \label{fig:time_evolution_of_action_ratois}
\end{figure}
Figure~\ref{fig:time_evolution_of_action_ratois} illustrates the time evolution of the action ratio for both sellers and buyers, each smoothed with a three-step moving average. 

\vskip0.5\baselineskip
\noindent \textbf{Seller behavior}

Early in the simulation (steps 0–10), sellers actively engaged in providing and updating data to populate the marketplace and improve their offerings. As the marketplace matured, the share of these productive actions declined sharply, and the dominant action shifted to \textit{do\_nothing}. By step 15, \textit{do\_nothing} accounted for more than 60\% of seller actions, increasing to nearly 80\% by step 40. This transition suggests that most sellers quickly stabilized their strategies, opting to maintain existing datasets rather than continually adjusting prices or refreshing data.

\vskip0.5\baselineskip
\noindent \textbf{Buyer behavior}

For buyers, the \textit{plan} action initially dominated as new entrants explored the market and formulated purchasing strategies. From around step 15, the shares of buying and analyzing began to grow steadily, while searching slowly diminished. This evolution reflects the transition from market exploration to regular purchasing and post-purchase analysis of datasets. By the final phase, \textit{buy} and \textit{analyze} accounted for more than 40\% of buyer actions, suggesting that buyers increasingly engaged in repeated or analytical usage of the dataset they had acquired.

\vskip0.5\baselineskip
\noindent \textbf{Linking micro to macro phenomena}

Table~\ref{tab:macro_micro_reproduction} summarizes for the results of each metric. The stabilization of seller behavior (dominance of \textit{do\_nothing}) effectively froze prices and the composition of available datasets. 
This supply-side stability reduced uncertainty for buyers, enabling them to shift from exploratory \textit{search} to persistent \textit{buy} and \textit{analyze} actions. As buyers repeatedly analyzed and re-purchased data with confidence in stable pricing, the overall transaction counts remained consistently high, producing the strong temporal autocorrelation observed at the market level. 
Moreover, repeated analysis of specific datasets fostered the late-stage concentration of demand in particular fields such as social media and environmental science. In this way, individual micro-decisions coalesced into macro-level regularities, where a stabilized supply landscape and sustained analytical demand reinforced each other to maintain and amplify market-wide trends.

\section{Conclusion}
We presented an LLM-MAS for simulating strategic and goal-oriented data marketplaces. Unlike conventional rule-based simulations, our framework enabled buyer and seller agents to reason in natural language, execute diverse strategies, and adapt to dynamic market conditions. Evaluation against real transaction data showed that LLM-MAS reproduced essential market features, such as scale-free transaction structures and buyer-seller networks, and captured the emergence of field-specific demand trends. These results highlight the potential of LLM-based agents to bridge micro-level decision making and macro-level market dynamics. These advances establish a foundation for more realistic market simulations and provide actionable insights for the design of future data marketplaces.

Despite these achievements, certain limitations remain. Discrepancies were observed in three metrics: (1) the number of transactions per buyer, (2) the number of transactions per (buyer, data) pair, and (3) the autocorrelation coefficient of transaction volumes over time. These gaps are likely influenced by the simplified modeling of participant entry. Future work should explore more sophisticated mechanisms, such as adaptive entry rules, incentive mechanisms, or LLM-based decision models, to achieve greater fidelity. Additionally, the current implementation restricts data to tabular formats and metadata-based analysis. While LLM hallucination is often cited as a threat to validity, its impact is limited in our setting. This is because agents are restricted to a finite set of actions, and metrics are computed from transactions rather than from textual claims. It is also important to refine the prompt design so that buyers better understand \textit{update\_frequency}. Specifically, when data with a high update frequency that aligns with their analytical objectives is discovered through \textit{search}, the prompt should be improved to enable actions such as planning a regular subscription in the next plan. Extending the framework to include richer data types (e.g., text, images) and enabling post-purchase analysis with actual samples would allow agents’ strategies to evolve from genuine analytical feedback, thereby enhancing realism and decision-support potential.

\section*{ACKNOWLEDGMENT}
This study was supported by the joint research project with Toyota Motor Corporation and JST PRESTO (JPMJPR2369).

\bibliographystyle{IEEEtran}
\bibliography{ref}

\begin{thebibliography}{10}
\providecommand{\url}[1]{#1}
\csname url@samestyle\endcsname
\providecommand{\newblock}{\relax}
\providecommand{\bibinfo}[2]{#2}
\providecommand{\BIBentrySTDinterwordspacing}{\spaceskip=0pt\relax}
\providecommand{\BIBentryALTinterwordstretchfactor}{4}
\providecommand{\BIBentryALTinterwordspacing}{\spaceskip=\fontdimen2\font plus
\BIBentryALTinterwordstretchfactor\fontdimen3\font minus \fontdimen4\font\relax}
\providecommand{\BIBforeignlanguage}[2]{{%
\expandafter\ifx\csname l@#1\endcsname\relax
\typeout{** WARNING: IEEEtran.bst: No hyphenation pattern has been}%
\typeout{** loaded for the language `#1'. Using the pattern for}%
\typeout{** the default language instead.}%
\else
\language=\csname l@#1\endcsname
\fi
#2}}
\providecommand{\BIBdecl}{\relax}
\BIBdecl

\bibitem{rusche2019data}
C.~Rusche, ``Data economy and antitrust regulation,'' \emph{Intereconomics}, vol.~54, no.~2, pp. 114--119, 2019.

\bibitem{balazinska2011data}
M.~Balazinska, B.~Howe, and D.~Suciu, ``Data markets in the cloud: An opportunity for the database community,'' \emph{Proceedings of the VLDB Endowment}, vol.~4, no.~12, pp. 1482--1485, 2011.

\bibitem{liang2018survey}
F.~Liang, W.~Yu, D.~An, Q.~Yang, X.~Fu, and W.~Zhao, ``A survey on big data market: Pricing, trading and protection,'' \emph{IEEE Access}, vol.~6, pp. 15\,132--15\,154, 2018.

\bibitem{fernandez2020data}
R.~C. Fernandez, P.~Subramaniam, and M.~J. Franklin, ``Data market platforms: trading data assets to solve data problems,'' \emph{Proceedings of the VLDB Endowment}, vol.~13, no.~12, pp. 1933--1947, 2020.

\bibitem{stahl2016classification}
F.~Stahl, F.~Schomm, G.~Vossen, and L.~Vomfell, ``A classification framework for data marketplaces,'' \emph{Vietnam Journal of Computer Science}, vol.~3, pp. 137--143, 2016.

\bibitem{schomm2013marketplaces}
F.~Schomm, F.~Stahl, and G.~Vossen, ``Marketplaces for data: an initial survey,'' \emph{ACM SIGMOD Record}, vol.~42, no.~1, pp. 15--26, 2013.

\bibitem{spiekermann2019data}
M.~Spiekermann, ``Data marketplaces: Trends and monetisation of data goods,'' \emph{Intereconomics}, vol.~54, no.~4, pp. 208--216, 2019.

\bibitem{abbas2021business}
A.~E. Abbas, W.~Agahari, M.~Van~de Ven, A.~Zuiderwijk, and M.~De~Reuver, ``Business data sharing through data marketplaces: A systematic literature review,'' \emph{Journal of Theoretical and Applied Electronic Commerce Research}, vol.~16, no.~7, pp. 3321--3339, 2021.

\bibitem{hayashi2020understanding}
T.~Hayashi and Y.~Ohsawa, ``Understanding the structural characteristics of data platforms using metadata and a network approach,'' \emph{IEEE Access}, vol.~8, pp. 35\,469--35\,481, 2020.

\bibitem{hayashi2022models}
T.~Hayashi, H.~Matsushima, H.~Sakaji, Y.~Fukami, and T.~Shimizu, ``Models of exchanged datasets and interactions of buyers in the data market: Toward multi-agent simulators for system design,'' \emph{Procedia Computer Science}, vol. 207, pp. 1695--1704, 2022.

\bibitem{sashihara2024impact}
J.~Sashihara and T.~Hayashi, ``Impact of buyer strategy and market size on data marketplace dynamics: A network-based simulation study,'' in \emph{2024 IEEE International Conference on Big Data (BigData)}.\hskip 1em plus 0.5em minus 0.4em\relax IEEE, 2024, pp. 6632--6641.

\bibitem{ijcai2024p874}
S.~Ren, Z.~Cui, R.~Song, Z.~Wang, and S.~Hu, ``Emergence of social norms in generative agent societies: Principles and architecture,'' in \emph{Proceedings of the Thirty-Third International Joint Conference on Artificial Intelligence, {IJCAI-24}}.\hskip 1em plus 0.5em minus 0.4em\relax International Joint Conferences on Artificial Intelligence Organization, 2024, pp. 7895--7903.

\bibitem{chiarella2002simulation}
C.~Chiarella and G.~Iori, ``A simulation analysis of the microstructure ofdouble auction markets,'' \emph{Quantitative finance}, vol.~2, no.~5, p. 346, 2002.

\bibitem{wang2024survey}
L.~Wang, C.~Ma, X.~Feng, Z.~Zhang, H.~Yang, J.~Zhang, Z.~Chen, J.~Tang, X.~Chen, Y.~Lin \emph{et~al.}, ``A survey on large language model based autonomous agents,'' \emph{Frontiers of Computer Science}, vol.~18, no.~6, p. 186345, 2024.

\bibitem{park2023generative}
J.~S. Park, J.~O'Brien, C.~J. Cai, M.~R. Morris, P.~Liang, and M.~S. Bernstein, ``Generative agents: Interactive simulacra of human behavior,'' in \emph{Proceedings of the 36th annual acm symposium on user interface software and technology}, 2023, pp. 1--22.

\bibitem{yao2023react}
S.~Yao, J.~Zhao, D.~Yu, N.~Du, I.~Shafran, K.~Narasimhan, and Y.~Cao, ``React: Synergizing reasoning and acting in language models,'' in \emph{International Conference on Learning Representations (ICLR)}, 2023.

\bibitem{bo2024reflective}
X.~Bo, Z.~Zhang, Q.~Dai, X.~Feng, L.~Wang, R.~Li, X.~Chen, and J.-R. Wen, ``Reflective multi-agent collaboration based on large language models,'' \emph{Advances in Neural Information Processing Systems}, vol.~37, pp. 138\,595--138\,631, 2024.

\bibitem{yang2025twinmarket}
Y.~Yang, Y.~Zhang, M.~Wu, K.~Zhang, Y.~Zhang, H.~Yu, Y.~Hu, and B.~Wang, ``Twinmarket: A scalable behavioral and socialsimulation for financial markets,'' \emph{arXiv preprint arXiv:2502.01506}, 2025.

\bibitem{piao2025agentsociety}
J.~Piao, Y.~Yan, J.~Zhang, N.~Li, J.~Yan, X.~Lan, Z.~Lu, Z.~Zheng, J.~Y. Wang, D.~Zhou \emph{et~al.}, ``Agentsociety: Large-scale simulation of llm-driven generative agents advances understanding of human behaviors and society,'' \emph{arXiv preprint arXiv:2502.08691}, 2025.

\bibitem{sakaji2020verification}
H.~Sakaji, T.~Hayashi, K.~Izumi, and Y.~Ohsawa, ``Verification of data similarity using metadata on a data exchange platform,'' in \emph{2020 IEEE International Conference on Big Data (Big Data)}.\hskip 1em plus 0.5em minus 0.4em\relax IEEE, 2020, pp. 4467--4474.

\end{thebibliography}

\newpage
\section*{Appendix}
We list the prompt templates for our simulation. 
\begin{figure}[h]
    \centering
    \begin{tcolorbox}[
    enhanced,
    colframe=black, 
    colback=white, 
    coltitle=black, 
    colbacktitle=white, 
    boxed title style={frame hidden}, 
    fonttitle=\bfseries,
    attach boxed title to top center={yshift=-4mm},
    ]
    \small{
    You are a data creator, who creates new data for ...\\
    
    \# Instructions\\
    - Field: \{field\}\\
    - Important Constraints: The data name must be unique ...\\
    
    \# Metadata content\\
    - data\_name: name of the data\\
    - description: description of the data content, which ...\\
    - columns: column names of the data (e.g. 'id,date,value')\\
    - tags: \{field\}\\
    - data\_price: price of the data\\
    - update\_frequency: update frequency of the data (high, ...)
    }
    \end{tcolorbox}
    \caption{Prompt for DataGenerator}
    \label{fig:prompt_for_datagenerator}
\end{figure}

\begin{figure}[h]
    \centering
    \begin{tcolorbox}[
    enhanced,
    colframe=black, 
    colback=white, 
    coltitle=black, 
    colbacktitle=white, 
    boxed title style={frame hidden}, 
    fonttitle=\bfseries\small,
    attach boxed title to top center={yshift=-4mm},
    ]
    \small{
    You are a data seller agent aiming to ...\\
    
    \# Available actions\\
    - `update\_data`: Update dynamic data by adding ...\\
    - `change\_price`: Lower the price ... or consider raising ...\\
    - `provide\_data': Create and provide new data ...\\
    - `do\_nothing': Observe ... and maintain ... \\
    - `exit\_market': If ... , you will withdraw from ...\\
    
    \# Notes \\
    - Use `do\_nothing' unless there is ...
    }
    \end{tcolorbox}
    \caption{System prompt for seller}
    \label{fig:system_prompt_for_seller}
\end{figure}
\begin{figure}[h]
    \centering
    \begin{tcolorbox}[
    enhanced,
    colframe=black, 
    colback=white, 
    coltitle=black, 
    colbacktitle=white, 
    boxed title style={frame hidden}, 
    fonttitle=\bfseries,
    attach boxed title to top center={yshift=-4mm},
    ]
    \small{
    You are a data analyst for ... follow the current market trend or focus on another area ... \\
    
    \# Current Market Trends (Top \{len(top\_trends)\}):\\
    \{trends\_str\}\\
    
    \# Your Field: \\
    \{field\}\\
    
    \# Note:\\
    - If there is no trend related to the field, ... \\
    
    \# Example 1: Field has trend related data\\
    - Field: marketing\\
    - Trend: 1. "Social media ...", 2. "Stock price data ... \\
    - Analysis Purpose: Analyze the correlation between ...\\
    }
    \end{tcolorbox}
    \caption{Prompt \underline{with} trends for GoalGenerator}
    \label{fig:prompt_with_trends_for_goalgenerator}
\end{figure}

\begin{figure}
    \centering
    \begin{tcolorbox}[
    enhanced,
    colframe=black, 
    colback=white, 
    coltitle=black, 
    colbacktitle=white, 
    boxed title style={frame hidden}, 
    fonttitle=\bfseries,
    attach boxed title to top center={yshift=-4mm},
    ]
    \small{
    You are a data analyst for ...\\
    
    \# Your Field: \\
    \{field\}\\
    
    \# Note:\\
    - The analysis purpose should be concise and specific.\\
    
    \# Example:\\
    ...
    }
    \end{tcolorbox}
    \caption{Prompt \underline{without} trends for GoalGenerator}
    \label{fig:prompt_without_trends_for_goalgenerator}
\end{figure}

\begin{figure}
    \centering
    \begin{tcolorbox}[
    enhanced,
    colframe=black, 
    colback=white, 
    coltitle=black, 
    colbacktitle=white, 
    boxed title style={frame hidden}, 
    fonttitle=\bfseries,
    attach boxed title to top center={yshift=-4mm},
    ]
    \small{
    \# Your final purpose\\
    \{self.purpose\}\\
    
    \# Your current state\\
    - Budget: \{self.budget\}\\
    - Owned data list: \{purchased\_data\_str\}\\
    
    \# Your action history\\
    \{action\_history\_str\}\\
    
    \# Last search results (Top candidates from the market DB)\\
    \{json.dumps(self.last\_search\_results)\}
    }
    \end{tcolorbox}
    \caption{User prompt for buyer}
    \label{fig:user_prompt_for_buyer}
\end{figure}

\begin{figure}
    \centering
    \begin{tcolorbox}[
    enhanced,
    colframe=black, 
    colback=white, 
    coltitle=black, 
    colbacktitle=white, 
    boxed title style={frame hidden}, 
    fonttitle=\bfseries,
    attach boxed title to top center={yshift=-4mm},
    ]
    \small{
    You are a data buyer agent in a data marketplace ... \\
    
    \# Available actions\\
    - `plan': Plan the next actions ... , or refine the purpose ... \\
    - `search': Search the marketplace for datasets that match ...\\
    - `buy': Purchase relevant data from the marketplace.\\
    - `analyze': Analyze the datasets ... to gain insights.\\
    - `do\_nothing': Intentionally take no action and ... \\
    - `exit\_market': Exit the marketplace and end the ...\\
    
    \# Guidelines for actions\\
    1. Begin by using `plan' to ...\\
    2. Based on the plan, ...\\
    3. If there is no suitable data ... , ... using `do\_nothing' ...\\
    4. Update or refine your plan, or ... , using `plan' ...\\
    5. Repeat this cycle until the final purpose is achieved or ...\\
    6. Always consider your budget and the datasets ...\\
    7. When you have achieved ... purpose ... `exit\_market' ...\\
    
    \# Notes\\
    - Even when dealing with the same data, there can be ...\\
    - If necessary, you should buy the same data ...
    }
    \end{tcolorbox}
    \caption{System prompt for buyer}
    \label{fig:system_prompt_for_buyer}
\end{figure}
\end{document}